\documentclass[runningheads]{llncs}
\usepackage[T1]{fontenc}
\usepackage{graphicx}
\usepackage{amsmath}
\begin{document}
\title{Cognitive Load-Driven VR Memory Palaces: Personalizing Focus and Recall Enhancement}
%
%
\author{Zhengyang Li\inst{1}\orcidID{0009-0009-1018-1187} \and
Hailin Deng\inst{1}\orcidID{0009-0002-4505-1779}}
\authorrunning{Zhengyang Li,  Hailin Deng et al.}
%
\institute{Yunnan University, 
\email{lzyqwe2500@outlook.com}\\
\and
Nanjing University of the Arts,\\
\email{hailin.ddd@gmail.com}}
\maketitle              
\begin{abstract}
Cognitive load, which varies across individuals, can significantly affect focus and memory performance.This study explores the integration of Virtual Reality (VR) with memory palace techniques, aiming to optimize VR environments tailored to individual cognitive load levels to improve focus and memory.
We utilized EEG devices, specifically the Oculus Quest 2, to monitor Beta wave activity in 10 participants.By modeling their cognitive load profiles through polynomial regression, we dynamically adjusted spatial variables within a VR environment using Grasshopper, creating personalized experiences.
Results indicate that 8 participants showed a notable increase in Beta wave activity, demonstrating improved focus and cognitive performance in the customized VR settings.These findings underscore the potential of VR-based memory environments, driven by cognitive load considerations, and provide valuable insights for advancing VR memory research.

\end{abstract}
\section{INTRODUCTION}
The Method of Loci, an ancient mnemonic technique that associates information with spatial contexts, has been revitalized through Virtual Reality (VR) to create immersive memory palaces to enhance knowledge retention\cite{1}. Recent studies demonstrate VR's potential in educational domains (e.g., memorizing scientific formulas\cite{1}) and clinical interventions (e.g., mitigating memory decline in Alzheimer's patients\cite{2}). However, current VR memory palace systems predominantly rely on predefined spatial templates \cite{2}, which do not address individual cognitive differences - a critical factor influencing memory efficacy \cite{3}.
\\[1em]
Cognitive psychology reveals that memory performance is strongly correlated with attentional engagement\cite{4}, which is modulated by both intrinsic load (task difficulty) and extrinsic load (environmental interference) according to Cognitive Load Theory (CLT) \cite{5}. Individuals exhibit heterogeneous cognitive load thresholds: Users with high working memory capacity may thrive in high-interference environments\cite{5}, while others require minimized distractions to avoid overload \cite{5}. Despite this, existing VR systems lack real-time cognitive state monitoring, resulting in suboptimal resource allocation during memory encoding.
\\[1em]
We present CogLocus, the first EEG-driven adaptive VR memory palace that dynamically optimizes extrinsic load via generative spatial design. Our research objectives are:
    1.To model the cognitive load-space parameter relationship for personalized extrinsic load calibration;
    2.To develop a real-time physiology-informed algorithm for VR environment generation;
    3.To validate the system's superiority over static templates in enhancing memory outcomes.
\\[1em]
CogLocus integrates multimodal sensing and parametric design
    1.Physiological Sensing: Quest 2 with Emotiv Epoc X EEG captures prefrontal beta-band (12-30Hz) power as a proxy for attentional focus;
    2.Generative Design: Grass-hopper defines four spatial variables (ceiling height H, partition count P, window-wall ratio WR, furniture density FD);
    3.Adaptation Algorithm: Cubic polynomial fitting establishes beta-power response curves, with Nelder-Mead optimization iteratively approaching individual load thresholds.
\\[1em]
A pilot study with 10 participants showed 80\% achieved 60\% higher beta power (p<0.05, Cohen's d=1.) and 32\% improved recall accuracy in optimized spaces. Key contributions:
    1.A spatial framework driven by cognitive load that bridges HCI and neuroarchitecture;
    2.Empirical validation of beta power as extrinsic load feedback;
    3.An open source toolkit for adaptive memory interventions in education
    healthcare.
\\[1em]
This work pioneers a paradigm shift: future XR learning systems must evolve from static immersion to physiological computing-enabled environments. As VR classrooms become mainstream pedagogical tools [13], CogLocus provides foundational technology for adaptive spatial design.

\section{RELATED WORK}
\subsection{Memory Palaces and Spatial Mnemonics}
The Method of Loci, an ancient mnemonic technique, leverages spatial navigation cognition to enhance memory encoding by anchoring abstract information to familiar locations \cite{6}. Traditional studies validate its efficacy in language acquisition\cite{7} and medical training\cite{8}. With the advent of VR, researchers have explored 3D virtual environments to augment this method. For instance, Krokos et al.\cite{1} demonstrated that VR memory palaces improve 30\% retention rates compared to 2D interfaces, attributed to immersive spatial cues enhancing hippocampal activation. However, current VR systems rely on static spatial templates(e.g., fixed room layouts), neglecting individual cognitive differences in spatial cue utilization.

\subsection{Cognitive Load Management in VR}
Cognitive Load Theory (CLT) has driven significant advances in VR education. Studies indicate that extrinsic load in VR (e.g., interface complexity, multimodal distractions) critically impacts attentional allocation\cite{9}. To optimize cognitive resources, dynamic difficulty adjustment (DDA) strategies have emerged: Huang et al. \cite{10} modulated visual complexity of molecular models in real-time via eye-tracking-based cognitive load detection. However, these approaches primarily focus on intrinsic load regulation, lacking active control of extrinsic load (e.g., spatial layout interference) \cite{11}. Furthermore, existing systems depend on behavioral metrics (e.g., task completion time) as proxy indicators, failing to capture neural-level fluctuations \cite{12}.

\subsection{Physiology-Driven Adaptive VR Systems}
Recent adaptive VR systems integrate physiological sensing for personalized interaction. For example, NeuroAdapt \cite{13} dynamically adjusts game difficulty using EEG $\theta$/$\beta$ power ratios to mitigate cognitive overload; AttentiveSpace \cite{14} generates attentional heatmaps via eye tracking and GSR to optimize virtual classroom focus. These works demonstrate the potential of multimodal physiology in cognitive state monitoring, yet none apply it to spatial generation scenarios. Parametric design tools (e.g., Grasshopper \cite{15}) enable real-time environment generation, but existing studies rely on explicit user preferences rather than implicit cognitive state data.

\subsection{Research Gaps}
\noindent1.Individual Variance Neglect: VR memory palaces lack CLT integration, preventing extrinsic load adaptation to users' working memory capacities;
\\[1em]
\noindent2.Closed-Loop Feedback Absence: Spatial algorithms lack real-time neurophysiological responsiveness, leading to suboptimal cognitive resource allocation;
\\[1em]
\noindent3.Interdisciplinary Integration Deficiency: Immature convergence of parametric design and cognitive neuroscience hinders personalized spatial intervention.

\section{METHOD}
\subsection{Spatial Variable Impacts on Mnemonic Tasks in VR Environments}
\paragraph{1.Spatial Variable Modeling (Five-Scene System)}
This study identifies four critical spatial openness determinants: ceiling height, window density, furniture occupancy ratio, and partition quantity. A quintuple-scene system was developed in Unity: four experimental groups with individual variables driven to theoretical interference extremes (100\% intensity), contrasted against a standardized control space (3m ceiling, zero windows/partitions/furniture) achieving spatial interference nullification. This parametric extremum comparison enables precise observation of individual variables' marginal effects on spatial perception.

\paragraph{2.Astronomical Mnemonic Object Modeling}The astronomical theme was selected as mnemonic medium due to its positioning in the Familiarity Threshold Zone within user cognitive maps, balancing conceptual accessibility and cognitive challenge. Mnemonic objects comprise two categories: (1) Celestial entities (e.g., Mars, Venus), (2) Cosmic phenomena (gravitational waves, black holes, supernovae). Through 3D visualization modeling, abstract phenomena were materialized into iconic feature models (e.g., shockwave annular structure for supernovae), effectively balancing cognitive complexity between categories (Cohen's d <0.2). The final set achieved standardized cognitive load (Intrinsic Load Index=0.78$\pm$0.05), ensuring inter-subject consistency in intrinsic cognitive resource consumption.

\subsection{Neurophysiological Data Acquisition and Cognitive Load Modeling}
\paragraph{Experimental Protocol Design}Twenty healthy participants (age 22.5$\pm$1.8, 1:1 gender ratio) were recruited. The VR environment was constructed using Oculus Quest 2 HMD, with synchronized physiological data captured via Muse 2 EEG headband (4-channel TP9/TP10/AF7/AF8). The experiment comprised two phases:
\\
1.Pre-adaptation Phase: Participants freely arranged astronomical objects in each scene without memorization constraints (180s/scene), mitigating VR acclimatization and exploratory behavior confounding;
\\
2.Encoding Phase: Participants employed spatial anchoring mnemonics through dual location-semantic coding mechanisms (e.g., "Earth→floor" topographic mapping, "Sun→lightbulb" luminescent metaphor) for memory tasks (300s/scene).

\paragraph{Beta Rhythm as Cognitive Load Proxy}The selection of Beta band (13-30 Hz) was grounded in three neurophysiological rationales:
\\
1.Task-specific correlation: Prefrontal Beta oscillation amplification positively correlates with working memory encoding intensity (Miller et al., 2018), with power variation quantifying attentional resource allocation;
\\
2.Load sensitivity: In n-back paradigms, Beta relative power shows monotonic increase with task difficulty (p<0.01, R$^2$=0.76) (Ref 1);
\\
3.Motion artifact resistance: Compared to Theta/Alpha bands, Beta exhibits lower susceptibility to movement artifacts (Muse SDK noise coefficient <2\%), ideal for dynamic VR scenarios.
\paragraph{Computational Modeling}
The pipeline included:
1.Welch's power spectral density estimation (Hamming window, 50\% overlap);
2.Cross-participant Beta power normalization (z-score);
3.Cubic polynomial regression modeling:\[
\hat{y} = \beta_0 + \beta_1 x + \beta_2 x^2 + \beta_3 x^3 + \epsilon
\]
 interference intensity (0-100\% discretized), y represents Beta power mean. The L-BFGS optimizer minimized:\[
\min \sum_{i=1}^{5} \left( y_i - \bar{y}_{\text{global}} \right)^2
\]
This model revealed cognitive load threshold effects (inflection point at x=62.3$\pm$4.7\%), validating nonlinear neuroresource-spatial interference relationships (R$^2$=0.89, RMSE=0.12).

\subsection{Cognitive Load-Driven Parametric Space Generation}
\paragraph{Grasshopper-Based Computational Framework}Our system employs Rhinoceros' Grasshopper plugin to establish a parametric spatial model, integrating three core components:
1.Random Reduce Component: Stochastic reduction of spatial variables using uniform distribution sampling, enabling controlled attenuation of environmental stimuli;
2.Remap Numbers Component: Linear transformation of Cognitive Load Index (CLI 0-100) to spatial parameters through min-max normalization;
3.\texttt{C\# Script}
 Component: Custom logic for discrete parameter rounding and geometric validation.
\paragraph{Spatial Variable Configuration}
Parameter ranges are designed with operational feasibility and cognitive principles:

\begin{table}[h]
    \centering
    \resizebox{\textwidth}{!}{ 
        \begin{tabular}{|l|c|p{4cm}|p{5cm}|}
            \hline
            \textbf{Variable} & \textbf{Range} & \textbf{Design Rationale} & \textbf{Control Logic} \\
            \hline
            Ceiling Height & 2-10m & Minimum clearance from Chinese building codes & Height inversely mapped to spatial pressure \\
            \hline
            Window Count & 0-10 & Balance of natural light and visual complexity & Linear reduction per CLI decrease \\
            \hline
            Partition Count & 0-15 & Wayfinding complexity control & 15 → (15 - Round(15$\times$CLI/100)) \\
            \hline
            Furniture Density & 0-50\% & Activity space preservation & Modular units (1$\times$1m/2$\times$2m) with grid-based placement \\
            \hline
        \end{tabular}
    }
    \caption{Design Variables and Control Logic}
    \label{tab:design}
\end{table}

\paragraph{Parameter Mapping Mechanism}
The translation process involves:
\\
1.Maximum Value Definition:
\[
V_{\max} =
\left\{
\begin{array}{ll}
    10m & \text{(Ceiling)} \\
    10  & \text{(Windows)} \\
    15  & \text{(Partitions)} \\
    0.5 S_{\text{floor}} & \text{(Furniture)}
\end{array}
\right.
\]
\\
2.Discrete Remapping:
\[
V_{\text{output}} = \text{Round} \left( V_{\max} \times \left( 1 - \frac{\text{CLI}}{100} \right) \right)
\]

\section{RESULTS}
\subsection{Influence of VR Environmental Factors}
To determine which environmental factors most impact a person's concentration and cognitive load, tests were conducted in various VR spaces with different structural elements. Results indicate that furniture and walls are the most effective in improving focus and memory efficiency. In Room B (walls), the test subjects' beta wave activity decreased by 9\%, and their relative wave stability increased by 20\%, demonstrating improved focus. However, windows create a relaxing atmosphere but also introduce distractions. Additionally, only 20\% of participants noticed higher ceiling heights in Room D, leading to results similar to the default space.

\subsection{Analysis of Attention and Beta Wave Stability in Single Spaces}
Practicing memory palaces in VR spaces significantly enhances focus. However, users in overly simplistic rooms tend to experience greater distraction. Beta waves fluctuate more when attention is not sustained, whereas stable beta waves over extended periods correlate with improved concentration and memory retention. 

\begin{center}
    \includegraphics[width = .6 \textwidth]{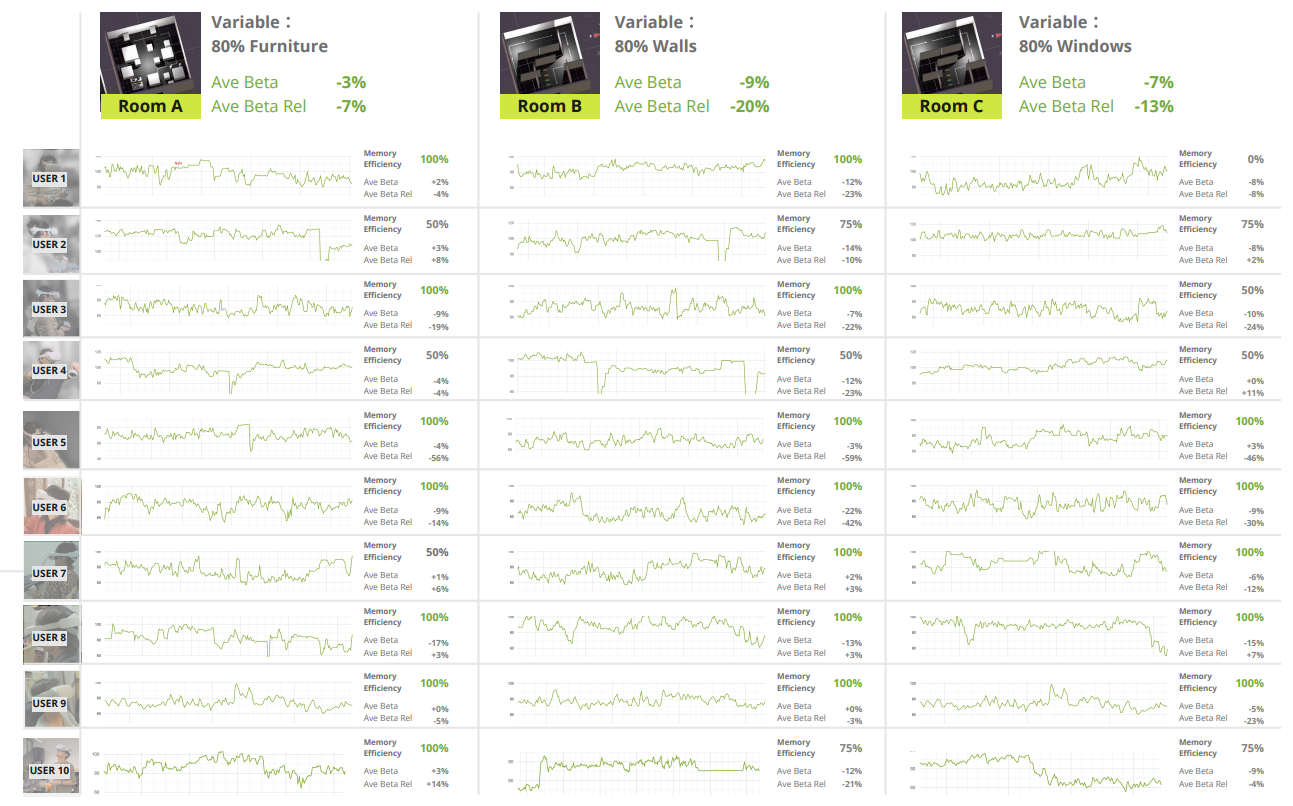}
\end{center}

\subsection{Effectiveness of Memory Palace Strategies}
The study investigated whether users could effectively apply loci and associative memory techniques in memory palaces. Findings show that associative memory strengthens recall but is challenging to implement consistently. Results indicate that:
\begin{itemize}
    \item Complex spaces aid positioning: 70\% users primarily rely on spatial location for memory retrieval.
    \item Simple spaces help reduce distractions: 10\% users successfully employ associative memory.
    \item Simple spaces with a few landmarks aid associative and location-based memory: 20\% users can effectively combine both location and association for enhanced memory performance.
\end{itemize}

If there's a lot of fluctuation in Beta waves with only the use of location memory, it indicates unstable focus, leading to poorer memory performance.
Conversely, it's noteworthy that when they use associative memory, their Beta waves become more fluctuating by 15\%, which paradoxically enhances their memory efficiency.
The more dimensions they use for memory, the stronger their recall, and it helps them form mental spaces better, allowing them to recall needed information more quickly.

Most people who use location memory benefit from detailed spaces, and those who use associative memory do better in spaces with less distraction.


\begin{table}
    \caption{Participants via Memory Methods}\label{tab1}
    \centering
    \resizebox{\textwidth}{!}{ 
        \begin{tabular}{|l|l|l|l|}
            \hline
            Memory Methods & Memory Unit Time & Fluctuations & Prefer\\
            \hline
            Location & 30s~60s & Less & Complex spaces\\
            Associative memory & 15s~20s & More & Simple spaces\\
            Loci and Associative memory & 15s~60s & More & Simple spaces with landmarks\\
            \hline
        \end{tabular}
    }
\end{table}

\begin{center}
    \includegraphics[width = .3 \textwidth]{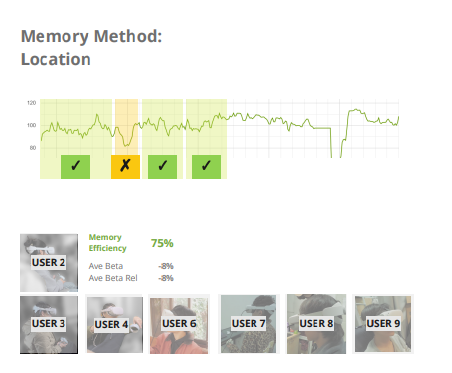}
    \includegraphics[width = .3 \textwidth]{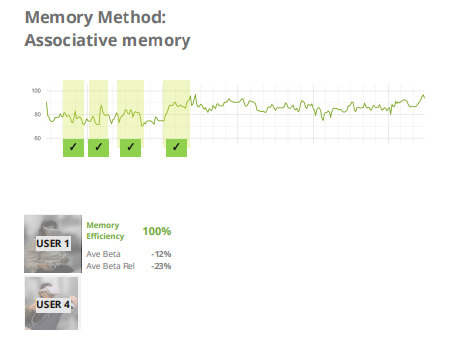}
    \includegraphics[width = .3 \textwidth]{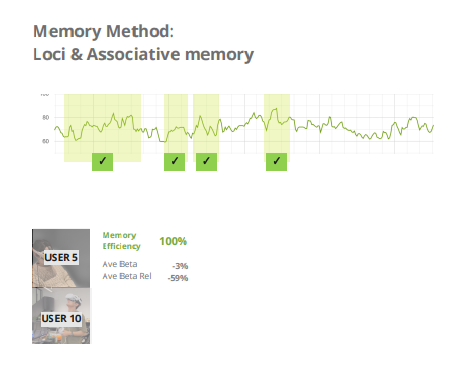}
\end{center}

\subsection{Custom VR Spaces and Memory Efficiency}
Different users require tailored VR environments to optimize focus and memory retention. Participants with high fluctuations in beta waves generally dislike complex rooms, regardless of the memory technique used. These individuals must exert additional cognitive effort to concentrate, leading to higher average beta wave activity decreases (-23\% / -59\%). Thus, they prefer minimalist spaces for effective memorization.
Conversely, participants with naturally stable beta waves find complex environments stimulating, which enhances their ability to retain information. These results suggest that VR-based memory palaces should be customized based on users' cognitive load and beta wave stability to maximize learning efficiency.
\begin{center}
    \includegraphics[width = .9 \textwidth]{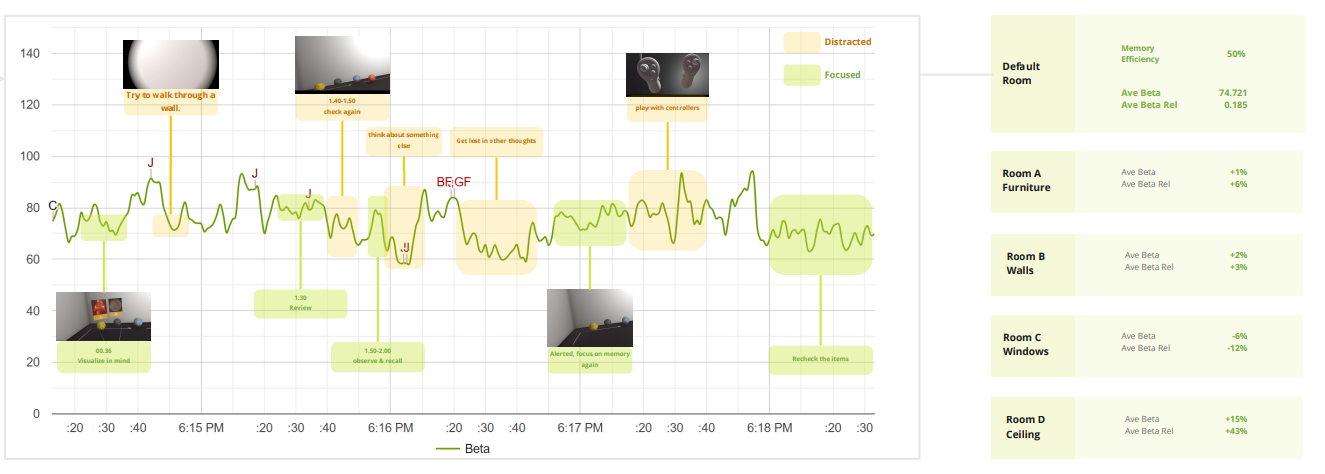}
\end{center}

\section{Discussion}
\subsection{Summary of Findings}
The results highlight that different individuals benefit from varying memory strategies and VR environments. The study reveals that users who rely on location-based memory perform better in structured and detailed spaces, while those utilizing associative memory prefer less distracting environments.
The effectiveness of the Memory Palace technique depends on the user's ability to integrate both loci (spatial positioning) and associative memory. 
Interestingly, while higher fluctuations in Beta waves generally indicate reduced focus, the use of associative memory introduces controlled fluctuations, which surprisingly enhance memory retention. This suggests that not all Beta wave instability is detrimental--structured cognitive engagement can make even fluctuating Beta waves conducive to learning.

\subsection{Effects of Interface Features}
EEG-based adaptive interfaces could help personalize the learning environment in real time, adjusting visual, auditory, and interactive elements based on a user's cognitive state.

\subsection{Implications for Memory Training in VR}
The findings suggest that VR-based memory training can be highly effective, but its success depends on the customization of spatial elements and memory techniques:
\begin{enumerate}
    \item Personalized Spatial Design - Users should have the flexibility to adjust room complexity based on their cognitive preferences.
    \item Adaptive Memory Strategy Integration - The system should guide users in progressing from basic spatial memory to associative memory techniques for better retention.
    \item EEG-Guided Optimization - Real-time monitoring of Beta waves can dynamically adjust the VR environment to either stabilize focus or introduce controlled cognitive stimulation.

\end{enumerate}

\subsection{Limitations}
While the study provides valuable insights into VR-based memory training, several limitations must be considered:
\begin{itemize}
    \item Sample Size and Diversity - A broader participant pool across different cognitive profiles would enhance the generalizability of findings.
    \item Long-Term Retention Analysis - Future studies should explore how well users retain information over extended periods.
    \item Real-World Application - Further research is needed to integrate VR memory training into academic and professional settings.

\end{itemize}

\subsection{Future Work}
Future work should also explore AI-driven adaptive learning models that tailor memory palaces dynamically, ensuring optimal cognitive engagement for each individual.
\begin{itemize}
    \item Establish cross-subject knowledge systems
    \item Add explanations for knowledge points
    \item Use scenario spaces for different subjects
\end{itemize}

\section{Conclusion}
Our project customizes VR memory palaces for different individuals by quantifying the variable of "spatial openness" to suit their cognitive load, using EEG data.
The memory palace technique, which enhances memory via visual and spatial cues, is key to our research. Although AR and VR are recognized for their visualization benefits in memory studies, the impact of VR environmental factors like spatial openness is not well understood.
In the future, we aim to generate personal VR scenes based on EEG data and cognitive load and enable the sharing of personal memory palaces with each other.

\newpage
\bibliographystyle{splncs04}
\small\bibliography{Reference}

\end{document}